# GW Explorer: A Beginner’s Guide — Developing a Computational Gravitational-Wave Outreach Curriculum for High School Students

Authors: Rachel Langgin (1), Bradlee Tejeda (2) and Carl-Johan Haster (1)

((1) University of Nevada, Las Vegas, (2) Vanderbilt University)

## Abstract

We present *GW Explorer: A Beginner’s Guide*, an outreach curriculum designed to introduce high school students to gravitational-wave (GW) astrophysics through interactive Python Jupyter notebooks. Most existing GW resources target beginning audiences and advanced students, leaving a gap at the pre-college level that we directly address. The curriculum integrates foundational physics with hands-on computation implemented through both self-directed and workshop-based instructional formats. In the self-directed format, students completed the curriculum independently on cloud-based platforms such as Google Colab. During the workshop format, students worked through the same activities under the guidance of University of Nevada, Las Vegas graduate student mentors. Topics span gravity, spacetime, GW sources, interferometric detection, and data analysis. An implementation in local high school classrooms informed the content and pacing, and survey results demonstrate gains in conceptual understanding and coding confidence. *GW Explorer* offers a scalable, open-access framework for authentic astrophysics research in the high school classroom.

## Additional keywords

# 1. Introduction

More than ten years have passed since the first direct detection of Gravitational waves (GWs) [LIGO Scientific Collaboration, 2016]. GWs are distortions in spacetime caused by the acceleration of massive objects that traverse the Universe at the speed of light. The most common sources of GWs are binary compact objects – colliding black holes or neutron star binaries [LIGO Scientific Collaboration, 2017], which induce a strain on spacetime, producing detectable signals that stretch and squeeze spacetime of the order of roughly one part in $10^{21}$ meters. Strain is the fractional change in distance between two points caused by a passing GW and is a measurable quantity.

The coalescence of compact binary systems provides a well-understood signature within the theory of gravitation given by General Relativity. As the two compact objects orbit, General Relativity predicts they lose orbital energy to gravitational radiation, causing the orbit to shrink and the objects to spiral inward, producing a chirp signal that grows in both frequency and amplitude over time. The modeled signals [LIGO Scientific Collaboration, 2016], called waveforms [Buonanno & Damour 2000; Ajith et al. 2007; Blanchet 2014], are patterns of stretching and squeezing of spacetime, described as strain versus time, and directly encode information about the source that created them. The search for GWs uses matched-filtering techniques [Usman et al., 2016; Cannon et al., 2021], in which noisy data from interferometric observatories, such as the two LIGO detectors, is correlated with a bank of template waveforms that describe all expected signals, separating the detector noise from the astrophysical signals of interest.

There exist many outreach efforts, as outlined in Middleton et al. 2024, focused on communicating GW discoveries to the wider public. An example specific to supporting formal classroom education include an Educator's Guide [Edeon STEM Learning, 2016] suitable for ages 12 and up which covers written GW background material and in-person demonstrations. As well as Einstein-First (www.einsteinianphysics.com) which has developed an eight-year curriculum *Eight Steps to Einstein's Universe* for students aged 7-8 and 15-16 years old and uses interactive learning methods to teach students about modern physics but currently does not contain a standalone module on GW astrophysics.

The Gravitational-Wave Open Science Center (GWOSC; https://gwosc.org) [LIGO Scientific Collaboration, Virgo Collaboration, KAGRA Collaboration, 2021, 2023, 2026, 2026] maintains a comprehensive collection of open-access educational resources, tutorials, and data products for the GW community. These materials include publicly released datasets from detectors within the LIGO-Virgo-KAGRA (LVK) Collaboration (the two LIGO sites in the U.S.A. [LIGO Scientific Collaboration, 2015], Virgo in Italy [Virgo Collaboration, 2015], and KAGRA in Japan [KAGRA Collaboration, 2021]), along with Python-based tutorials covering data access, signal processing, and basic analysis workflows.

The tutorials and workshops function well at the advanced undergraduate and graduate level. However, their mathematical and computational demands create real barriers for pre-college students [Farr et al., 2012] or those with limited programming backgrounds.

There is a clear need for structured introductory materials that reduce barriers to entry without sacrificing the core physical concepts and data-driven character of GW astrophysics that this project aims to address.

In this insight, we detail *GW Explorer: A Beginner's Guide* and the platform. We first review the computational notebook design and the concepts taught therein. Next, we provide reflections from outreach initiatives including both a classroom implementation and independent learning session. We conclude with implications upon the use of computational interactive activities for learning GW astrophysics.

## 1.1 GW Explorer: A Beginner's Guide

The *GW Explorer: A Beginner's Guide* (*GW Explorer*) curriculum introduces high school students to compact objects, GWs, and interferometric detection through a structured sequence of Python-based Jupyter notebooks [Kluyver et al., 2016; Barba et al., 2019]. Each notebook weaves together narrative explanations, interactive visualizations, guided questions, and executable code, allowing students to explore concepts such as binary orbits and waveform generation directly within a computational environment. Short exercises, targeted coding tasks, and visualization tools support self-paced exploration while maintaining a coherent pedagogical progression.

To reduce barriers to entry, the curriculum opens with an introductory Python module covering variables, loops, functions, and data plotting – enough that students with no prior coding experience can engage with the scientific content immediately. We designed workshops to pair students with graduate volunteers who serve as near-peer mentors, offering real-time guidance through unfamiliar computational and physical terrain. The notebooks are also designed and were tested to work as standalone, self-guided instructional materials.

All current *GW Explorer* notebooks and instructional materials are publicly available at https://github.com/rlanggin/GW_Explorer_A_Beginners_Guide (mirrored at https://rachellanggin.com/tutorials). Overall, *GW Explorer* is designed to lower barriers to participation in astrophysics by pairing accessible computational tools with authentic scientific content. In doing so, the project provides early exposure to research-grade concepts, strengthens computational literacy, and broadens participation in physics both within the Las Vegas community and beyond.

### 1.1.1 Computational Notebook Design

The GW Explorer curriculum is built as a progressive sequence of computational Jupyter notebooks that guide students from foundational GW physics to simplified yet authentic data analysis workflows. Each notebook combines narrative explanations, executable Python code, guided exercises, and interactive plotting in a unified environment. The scaffolded structure is intentional as students build enough conceptual and computational footing in early notebooks to engage meaningfully with GW data by the end of the sequence.

*Notebook 1: Introduction to Gravitational Waves* builds the physical foundations: gravity, spacetime, compact objects, and GW generation. Students explore how accelerating massive bodies perturb spacetime through conceptual discussion, visual demonstrations, and introductory coding tasks. A Python tutorial covering variables, loops, functions, arrays, and plotting is embedded directly in the notebook. A static snapshot of an interactive visualization is shown in Figure 1. The interactive plots are designed to develop intuition for compact-object binary systems, such as binary black holes and neutron stars, and their orbital evolution within General Relativity, under GW emission, while simultaneously introducing students to computational workflows.

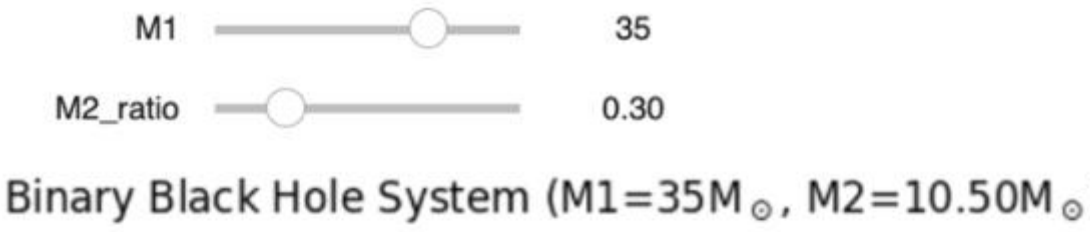

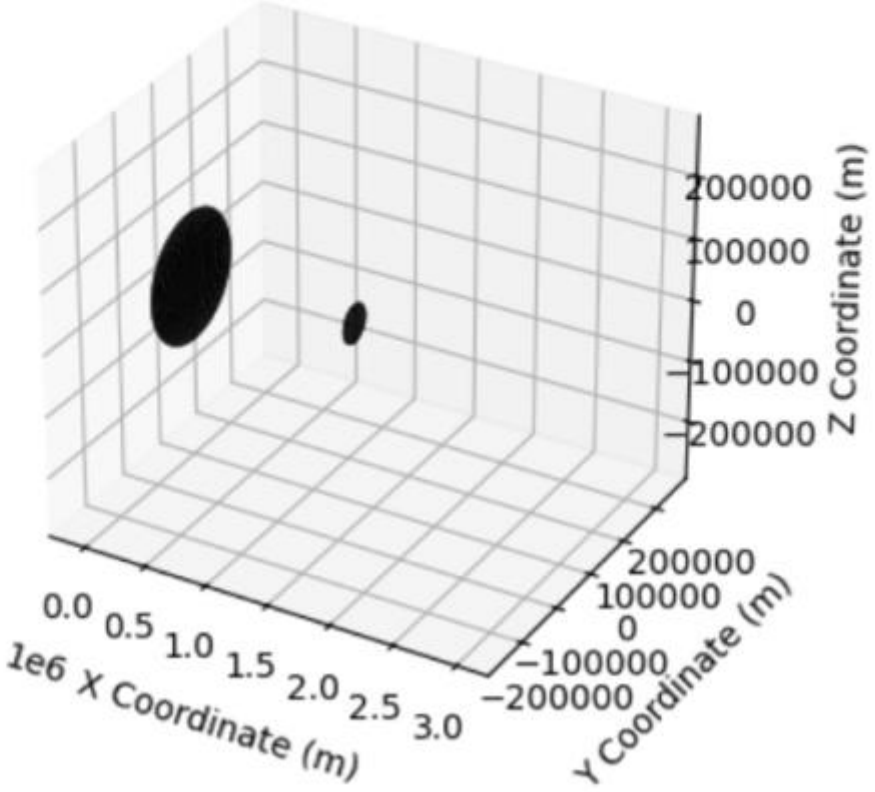

Figure 1. Example output from *GW Explorer* computational Notebook 1 introduces the students to compact binary coalescences through sliders and interactive Python widgets. This is a static snapshot of the interactive widget; in the notebook, students drag sliders to adjust component masses, which update in real time as they do so, to display binary mass-ratio configurations.

*Notebook 2: Simulating a Binary Merger* focuses on GW waveform morphology and the behavior of compact binary inspirals [Buonanno et al. 2007; Hannam et al. 2014; Bohé et al. 2016]. Students investigate how binary parameters – individual binary component masses, the ratio of the binary component masses, the shape and orientation of the binary orbit, with spin effects introduced later in a bonus module – shape GW signals through waveform simulations generated directly in the notebooks. Real-time waveform plotting and template matching activities allow students to visualize waveform amplitude and time evolution during inspiral and merger, as shown in Figure 2. By adjusting the binary's physical parameters and observing the waveform in real time, students build a genuine intuition for chirp morphology, detector strain, and the physical meaning of GW observables.

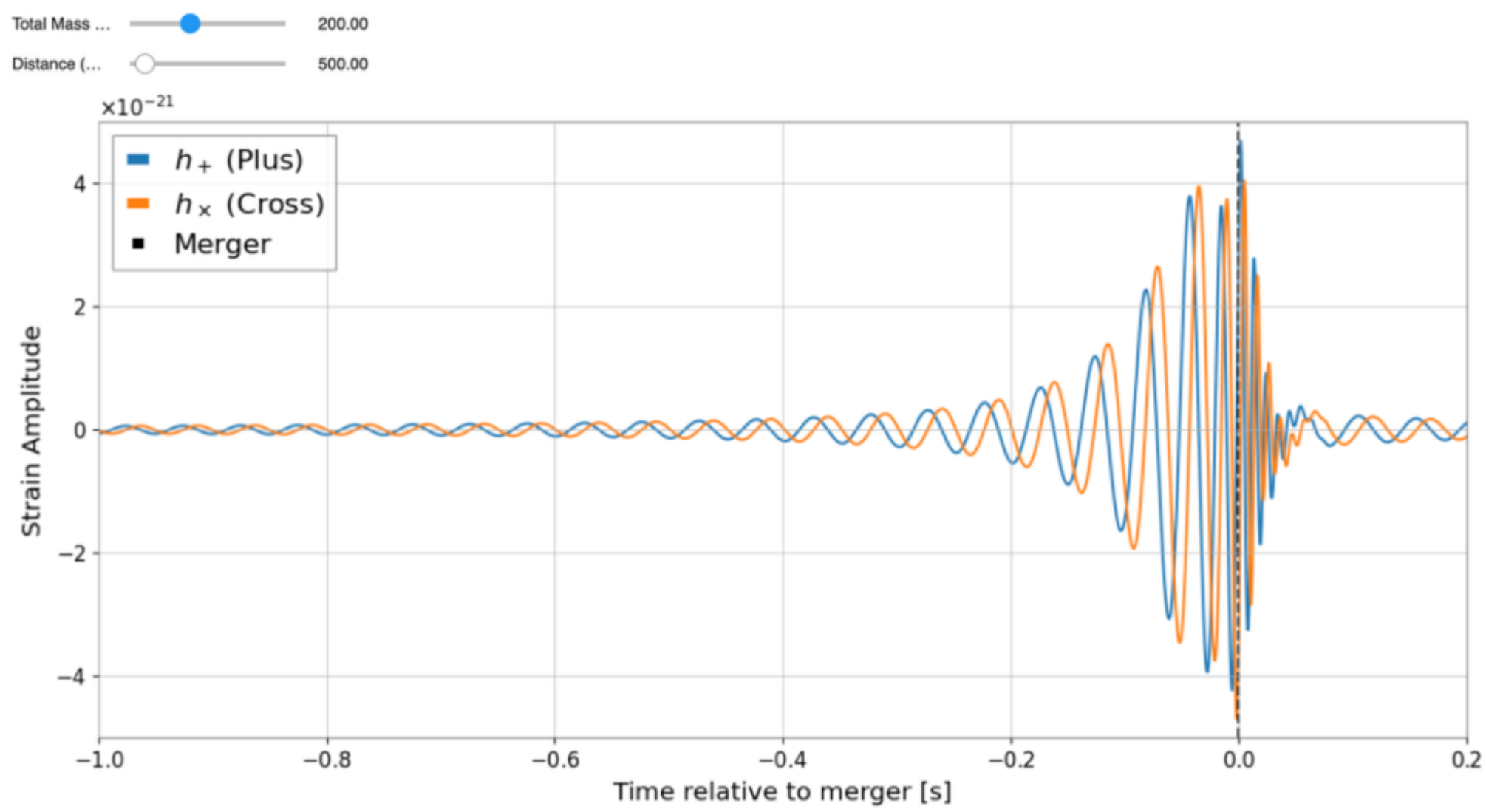

Figure 2. Example output from *GW Explorer* computational Notebook 1 introduces the students to GW waveform through sliders and interactive Python widgets. This is a static snapshot of the interactive widget; in the notebook, students drag sliders corresponding to ranges of total mass and luminosity distance, which update in real time as they adjust parameters, so the students visualize how different variables affect the amplitude and phase of the signal.

*Notebook 3: Parameter Estimation Basics* introduces students to simplified Bayesian inference concepts commonly used in modern GW astrophysics [Thrane & Talbot 2019]. Building on the waveform simulations from earlier notebooks, students learn how observational data are compared with theoretical models using likelihood evaluation and parameter estimation techniques, as shown in Figure 3. Guided exercises introduce probability distributions, model comparison, posterior sampling, and population statistics. The goal is to give students a working conceptual handle on the inference frameworks that underpin modern GW astronomy, at a simplified level.

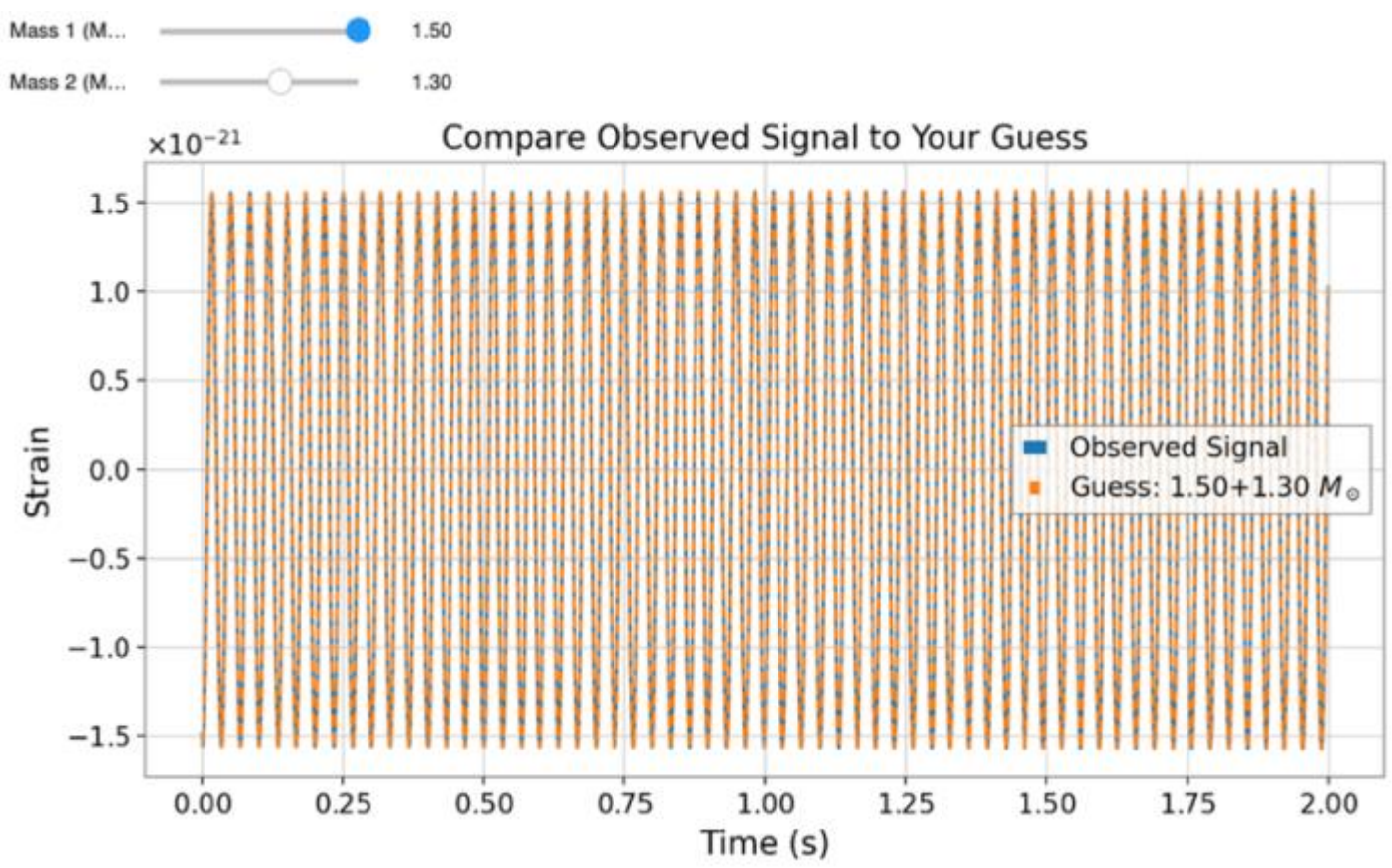

Figure 3. Example output from *GW Explorer* computational notebook 3 introduces the students to Bayesian likelihood analysis through sliders and interactive Python widgets. This is a static snapshot of the interactive widget; in the notebook, students drag sliders and watch the likelihood/match value update in real time as they adjust parameters.

# 2. Outreach Initiative

To evaluate the effectiveness of the *GW Explorer* curriculum, we conducted a pilot outreach program in collaboration with local Las Vegas high school classrooms. These sessions were designed to assess both conceptual understanding of GW topics and student engagement with computational activities.

Each workshop opened with a brief presentation on GW astrophysics, followed by a demonstration of a tabletop Michelson interferometer and guided interaction with the Jupyter notebooks. Students worked through selected modules with support from University of Nevada, Las Vegas (UNLV) graduate mentors who received a short training session covering the notebook content, common student misconceptions about GW physics, and troubleshooting for typical coding errors, and who provided real-time feedback and individualized help. Observations from these sessions, together with student survey responses, drove iterative refinements to the curriculum, including adjustments to content clarity, pacing, and coding difficulty that carried forward into later implementations.

## 2.1 Las Vegas Academy of the Performing Arts

We conducted an after-school workshop at Las Vegas Academy of the Performing Arts to evaluate the *GW Explore*r curriculum. As a school with a primary emphasis on the performing arts, it provided an opportunity to assess student engagement with computational astrophysics activities in a nontraditional STEM-focused setting. The workshop was hosted through the school's Mathematics Club, resulting in participation from students with a preexisting interest in mathematics and related STEM subjects.

Figure 4 highlights student participation led by UNLV graduate student mentors during the Las Vegas Academy *GW Explorer* workshop.

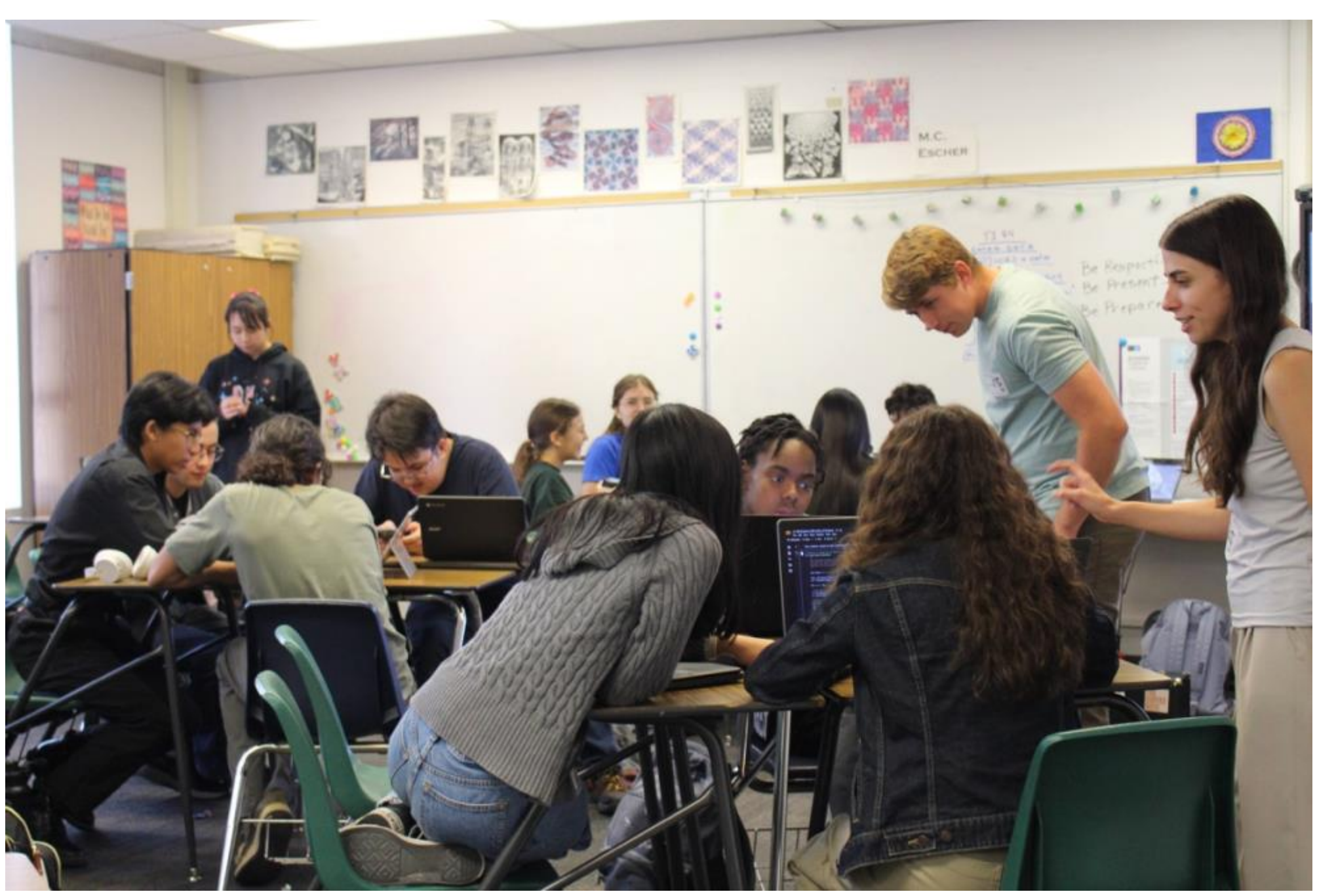

Figure 4. Las Vegas Academy of the Arts high-school students working through *GW Explorer* computational notebook activities with guidance from University of Nevada, Las Vegas graduate mentors.

The students received a short lecture on an introduction to physics, astronomy, and coding that reviewed conceptual definitions that the students would come across during completion of the notebooks.

### 2.1.1 Pre-Survey Results

We administered a pre-survey (N = 17) to assess students' prior exposure to physics, astronomy, and computation. Table 1 shows that after the short lecture, most students had some familiarity with basic astronomy, including black holes, the solar system, and galaxies, but relatively few encountered neutron stars or GWs.

Coding experience was mixed: 59% of respondents reported some prior exposure and 65% reported at least baseline confidence. The mathematics experience reported strong confidence in algebra (94%), while comfort with calculus was lower (59%).

Most students identified informal media sources – YouTube, social media apps, and television – as their primary exposure to astronomy, reflecting limited formal instruction on these topics. Notably, 59% expressed interest in pursuing a STEM-related career, suggesting a genuine motivation to engage with the workshop activities.

| Pre-Survey Item: | Percentage (%) |
|---|---|
| Prior coding experience | 59% |
| Prior coding self-confidence | 65% |
| Familiar with black holes | 88% |
| Familiar with neutron stars | 41% |
| Comfortable with algebra | 94% |
| Comfortable with calculus | 59% |
| Primary exposure to astronomy through media | 59% |
| Pursuing STEM career | 59% |

Table 1. High-school pre-survey summary of students' prior exposure to STEM topics reported by 17 survey respondents.

### 2.1.2 Post-Survey Results

A post-survey (N = 9) assessed understanding and self-reported gains following the workshop. The drop from 17 pre-survey respondents to 9 post-survey respondents reflects time constraints at the end of the after-school session – several students had to leave early to catch transportation.

As shown in Table 2, 89% of post-survey respondents correctly identified GWs as spacetime perturbations. Fewer (44%) identified binary mergers as the primary GW source – a reasonable outcome given the students only covered Notebook 1, and one that points to where and how the curriculum can be strengthened.

A majority (78%) reported increased interest in astrophysics. Coding confidence improved noticeably: 67% of respondents rated themselves at an intermediate level after completing the later notebooks, compared to just 33% who reported beginner-level confidence after the introductory module.

Every student rated the notebooks as useful (100%) and expressed interest in additional *GW Explorer* materials (100%). Qualitative feedback consistently pointed to the interactive elements – especially real-time plotting, loops, and function manipulation – as the features most responsible for building comfort with computation, even among students who had never coded before.

| Post-Survey Item: | Percentage (%) |
|---|---|
| Understood GWs as spacetime perturbations | 89% |
| Identified binary mergers as GW sources | 44% |
| Increased interest in astrophysics | 78% |
| Beginner confidence in coding (Notebook 1) | 33% |
| Intermediate confidence in coding (Notebooks 2 and 3) | 67% |
| Usefulness of *GW Explorer* Notebook 1 | 100% |
| Desired additional *GW Explorer* Notebooks | 100% |

Table 2. High-school post-survey indicators of conceptual understanding and self-reported gains reported by 9 survey respondents.

### 2.1.3 Implications

Students consistently pointed to interactive plotting as the most helpful element for understanding waveform morphology and detector response. Taken together, the reported gains in astronomical interest and coding confidence suggest that pairing conceptual instruction with hands-on computational activities can meaningfully support both engagement and skill development – even in a single after-school session.

These results support the core pedagogy behind *GW Explorer,* which holds that executable code, guided exercises, and real-time visualization together create an effective framework for introducing GW concepts at the high school level. Given the small sample, the findings are preliminary – but they indicate that the *GW Explorer* notebooks can meaningfully expose students to computational astrophysics, build foundational coding skills, and engage learners across a wide range of prior experience. The combination of interactive design and mentor support helps bridge preparation gaps and keeps students in an active, exploratory mode.

Beyond the curriculum itself, students also arrived with questions about STEM and academic career pathways. UNLV graduate mentors were well-positioned to answer those questions, share their own trajectories, and connect students directly with UNLV college application materials. This outreach dimension was just as valuable as the notebooks and resource exposure.

## 2.2 UNLV Undergraduate Physics Students

To probe the scalability of the *GW Explorer* curriculum beyond the high school context, we ran a pilot with undergraduate students at UNLV as a self-guided learning mode. The cohort spanned mechanical engineering, biological sciences, computer science, construction management, medical imaging, and related STEM disciplines. All students in the undergraduate cohort were enrolled in an introductory physics course and had experience working in a laboratory setting. However, this group was heterogeneous in terms of both mathematical preparation and programming background.

### 2.2.1 Pre-Survey Results

A pre-assessment (N = 20) was used to evaluate baseline preparation in physics, mathematics, and computing, as well as conceptual familiarity with GW topics. Table 3 summarizes prior academic preparation and computational experience. It should be noted that since this cohort of undergraduate students completed the program via self-instruction that they did not receive a short beforehand.

Most participants had completed at least introductory university physics and calculus through level I or II, with a few students also completing multivariable calculus, differential equations, or AP-level equivalents. Programming backgrounds varied considerably: some arrived with no coding experience, while others had worked in Python, C++, or Java, yet the majority still self-identified as beginner-level programmers before engaging with the curriculum.

| Pre-Survey Item: | Percentage (%) |
| --- | --- |
| Completed at least introductory university physics | 100% |
| Completed Calculus I or higher | 100% |
| Completed Calculus II or higher | 75% |
| Prior programming experience | 40% |
| Self-reported beginner coding level | 85% |
| Prior exposure to GW or LIGO concepts | 30% |
| Familiarity with black holes | 95% |
| Familiarity with neutron stars | 60% |

Table 3. Undergraduate pre-assessment of academic background and computational experience (N = 20).

Conceptually, students generally linked black holes to high-density gravitational collapse. Still, descriptions of GW sources were much less precise – most responses gestured at “cosmic collisions” or general orbital motion without invoking compact binaries or spacetime perturbations. This baseline confirmed that the curriculum was starting from the right place.

### 2.2.2 Post-Survey Results

A post-assessment (N = 18) was administered after completion of the *GW Explorer* notebooks. Two participants did not complete the full curriculum (N = 20 to 18). Table 4 summarizes post-completion conceptual understanding and self-reported outcomes.

| Post-Survey Item: | Percentage (%) |
|---|---|
| Correctly identified GWs as spacetime perturbations | 94% |
| Identified compact binary mergers as GW sources | 78% |
| Reported increased understanding of GW astrophysics | 89% |
| Reported increased coding confidence | 83% |
| Transitioned from beginner to intermediate coding confidence | 61% |
| Reported increased interest in astrophysics | 72% |
| Found notebooks useful or very useful | 100% |
| Requested additional modules or advanced content | 89% |

Table 4. Undergraduate post-assessment of conceptual understanding and self-reported outcomes (N = 18).

Reaffirming the high school results, qualitative responses singled out interactive waveform plotting and parameter exploration as the most influential components. Students repeatedly noted that real-time plotting and executable code improved their intuition for waveform morphology and detector response in a way that passive instruction had not.

### 2.2.3 Implications

Across the undergraduate cohort, results indicate measurable gains in both conceptual understanding and computational confidence following engagement with the *GW Explorer* curriculum. Notably, improvements were observed even among students with prior programming experience, who primarily reported gains in scientific interpretation of code rather than syntax comprehension. These findings suggest that the curriculum targets the translation of physical intuition into computational practice, which conventional instruction leaves largely undeveloped.

The survey data make a strong case that interactive computational notebooks effectively bridge the gap between theoretical instruction and data-driven GW analysis. The scaffolded progression and visualization-driven design proved robust across heterogeneous backgrounds, including the more mathematically prepared undergraduate cohort, indicating that the curriculum's pedagogical value is not contingent on prior knowledge gaps alone but reflects a deeper restructuring of how students engage with scientific computation.

## 3. Conclusion

This work presents *GW Explorer: A Beginner's Guide*, a computational outreach curriculum that provides high school students with access to GW astrophysics through interactive Jupyter notebooks in Python. By integrating conceptual physics instruction and interactive scientific coding across both mentor-supported workshops and self-guided learning modes, the curriculum lowers barriers to participation in computational GW astrophysics while keeping students engaged with authentic scientific ideas and workflows.

Unlike many traditional outreach approaches that emphasize passive learning or demonstrations, *GW Explorer* engages students with the computational and physical foundations of modern astrophysics through direct interaction with executable code and data-driven visualization. Students explore waveform morphology, compact binary systems, detector response, and parameter estimation directly within notebook environments that reflect tutorial versions of contemporary scientific analysis pipelines. The curriculum therefore does more than introduce astronomy content; it exposes students to the computational reasoning practices that increasingly define modern scientific research.

Pilot implementations with both high school and undergraduate cohorts demonstrate that students from diverse educational backgrounds can successfully engage with foundational GW concepts and computational tools. Survey results document gains in conceptual understanding of GWs as spacetime perturbations, compact binary mergers, and interferometric detection, alongside improvements in coding confidence and sustained interest in astrophysics. Across both cohorts, students point to real-time visualization and parameter manipulation as the most valuable components. This consistently suggests that interactive computational environments help close the gap between abstract physical concepts and intuitive understanding.

Critically, these outcomes were observed even for students with minimal prior programming experience. The introductory Python module, combined with near-peer mentor support [Vanasupa et al. 2014], allowed students to engage meaningfully with scientific computing regardless of where they started. This is perhaps the most important result: it suggests that computational astrophysics can be introduced effectively much earlier in students' educational trajectories than is typically assumed. This may also reflect the modern scientific research environment, as large-scale data analyses dominate workflows and computational thinking and proficiency become a priority [Astor et al. 2026].

More broadly, these results reinforce the case for computation-centered pedagogy in physics education – one in which coding, visualization, and data interaction are treated as core scientific literacy skills rather than as supplementary technical electives [Sengupta et al. 2013]. Modern GW astronomy is fundamentally computational, built on large-scale data analysis, simulation, and statistical inference. Access to authentic versions of those practices has historically been out of reach for pre-college students. By leveraging browser-based notebooks and open scientific resources, *GW Explorer* demonstrates that this need not be the case.

## 3.1 Discussion

The results of this work suggest that interactive computational notebooks provide an effective and scalable framework for introducing complex astrophysical concepts while simultaneously building computational literacy. In many introductory STEM settings, coding is treated as a supplemental skill to scientific instruction. In *GW Explorer*, the curriculum integrates programming directly into conceptual exploration, allowing students to manipulate physical parameters, generate visual outputs, and investigate scientific behavior through hands-on experimentation. This design reflects the increasingly computational nature of modern scientific practice, where simulation, visualization, and data analysis are central components of research workflows.

One of the clearest findings across both cohorts was the emphasis on visualization-driven learning. Students consistently singled out waveform plotting and detector-response activities as decisive for their understanding of binary mergers and GW propagation. The immediate visual feedback from executable code appears to strengthen conceptual intuition that static instruction cannot replicate, a finding consistent with broader evidence on interactive visualization in physics education [Perkins et al. 2006, Banda et al. 2021].

The near-peer mentor structure for the high school workshop was a meaningful complement to the notebooks themselves. UNLV graduate students provided individualized troubleshooting, modeled active scientific practice, and helped reduce the intimidation factor of unfamiliar computational tools. This kind of support seems to be key in encouraging the exploratory, experimental learning setting the notebooks are designed to cultivate.

Several limitations should be considered when interpreting these results. Sample sizes in both pilots were modest, post-survey completion was incomplete, and our assessment relied primarily on self-reported confidence and conceptual recognition rather than standardized measures. The workshop format also introduces variability in mentor interactions and classroom pacing that can be difficult to maintain compared to self-guided instruction. Future work should incorporate repeated assessment and, where feasible, controlled instructional comparisons to characterize long-term learning outcomes better.

## 3.2 Outlook

Future development of *GW Explorer* is expanding into a modular, scalable computational astrophysics education framework. Planned notebook modules will cover source localization, noise characterization, multi-messenger astronomy, and introductory machine learning for GW analysis. These additions will further align the curriculum with contemporary research practice while keeping it accessible to novice learners.

A primary long-term objective remains full integration within the GWOSC educational ecosystem. That integration would improve discoverability, support broader adoption, and position *GW Explorer* within established open-science infrastructure. This would also enable adaptation by classrooms, outreach programs, and independent learners well beyond Las Vegas to find, use, and adapt the materials.

Future development will also prioritize translating GW Explorer into Spanish, the most widely spoken non-English language in Southern Nevada homes (22.41% of Clark County residents age 5+; Claritas, 2026). This translation will be conducted by an undergraduate student at UNLV, furthering the project's commitment to accessible, community-grounded science education.

As GW astronomy continues to expand through next-generation observatories [Evans et al. 2022] and multi-messenger discovery, the need for accessible and computationally grounded educational resources will grow. *GW Explorer* demonstrates that frontier scientific topics can be introduced meaningfully at the pre-college and early undergraduate levels through interactive, research-connected learning environments. More broadly, this work suggests that open-access computation-centered pedagogy is an effective model for expanding scientific literacy, strengthening computational fluency, and opening access to modern astrophysics research.

# Acknowledgments

We thank the Las Vegas Academy of the Performing Arts and its mathematics teacher, Teresa Prezgay, for hosting the GW Explorer after-school workshop. We also acknowledge the UNLV graduate student mentors: Leah Green, Ted Johnson, and Madeline Overton for their assistance during workshop activities and student mentoring. Finally, we thank Marios Kalomenopoulos for guidance, initial notebook review, and testing of the GW Explorer curriculum. RL acknowledges support from a 2025 Forum on Outreach and Engaging the Public (FOEP) Mini-Grant from the American Physical Society. CJH is supported by the Nevada Center for Astrophysics, NASA 80NSSC23M0104. RL and CJH also acknowledge the support from the National Science Foundation through Award PHY-2409727.

This research has made use of data or software obtained from the Gravitational Wave Open Science Center (gwosc.org), a service of the LIGO Scientific Collaboration, the Virgo Collaboration, and KAGRA. This material is based upon work supported by NSF's LIGO Laboratory which is a major facility fully funded by the National Science Foundation, as well as the Science and Technology Facilities Council (STFC) of the United Kingdom, the Max-Planck-Society (MPS), and the State of Niedersachsen/Germany for support of the construction of Advanced LIGO and construction and operation of the GEO600 detector. Additional support for Advanced LIGO was provided by the Australian Research Council. Virgo is funded, through the European Gravitational Observatory (EGO), by the French Centre National de Recherche Scientifique (CNRS), the Italian Istituto Nazionale di Fisica Nucleare (INFN) and the Dutch Nikhef, with contributions by institutions from Belgium, Germany, Greece, Hungary, Ireland, Japan, Monaco, Poland, Portugal, Spain. KAGRA is supported by Ministry of Education, Culture, Sports, Science and Technology (MEXT), Japan Society for the Promotion of Science (JSPS) in Japan; National Research Foundation (NRF) and Ministry of Science and ICT (MSIT) in Korea; Academia Sinica (AS) and National Science and Technology Council (NSTC) in Taiwan.